\documentclass[english]{article}
\usepackage[T1]{fontenc}
\usepackage[latin9]{inputenc}

\usepackage{amssymb}
\usepackage{graphics}
\usepackage{pstricks,psfrag}
\usepackage{amsmath}
\usepackage{subfigure}
\usepackage{vmargin}
\usepackage{epsfig}

\usepackage{babel}

\begin{document}

\begin{titlepage}

\title{\bf Universality of Strength of Yukawa Couplings, Quark Singlets and Strength of CP Violation}

\vskip 1cm

\author{G.C. Branco\footnote{e-mail: {\tt gbranco@ist.utl.pt}}, 
H.R. Cola\c{c}o Ferreira\footnote{e-mail: {\tt hugoferreira@cftp.ist.utl.pt}}, 
A.G. Hessler\footnote{e-mail: {\tt ahessler@cftp.ist.utl.pt}} and 
J.I. Silva-Marcos\footnote{e-mail: {\tt juca@cftp.ist.utl.pt}} \\ \\
{\it  CFTP, Departamento de F\'{\i}sica} \\
{\it  Instituto Superior T\'ecnico,
Avenida Rovisco Pais, 1} \\
{\it 1049-001 Lisboa, Portugal}}
\maketitle

\vskip 1cm

\begin{abstract}
We analyse the strength of CP violation in an extension of the standard model with an extra
$Q=-1/3$ vector-like singlet quark, in the framework of the hypothesis of universality of strength of Yukawa couplings
connecting standard quarks. We show that the correct pattern of quark masses and mixing
can be obtained, including the observed strength of CP violation.
\end{abstract}

\end{titlepage}

\section{\protect\bigskip Introduction}

At present, there is no deep understanding of the observed pattern
of fermion masses and mixing. In the framework of the Standard Model
(SM), this pattern can be accommodated by adjusting the value of the
Yukawa couplings, which is allowed, since the gauge symmetry does
not constrain the flavour structure of the these couplings. There
have many attempts at finding a rationale for the pattern of fermion
masses and mixing \cite{frit}.

An interesting suggestion for the flavour pattern of Yukawa couplings
is the hypothesis of universal strength of Yukawa couplings (USY)
\cite{usy-first}, which assumes that the flavour dependence is all
contained in their phases. The USY hypothesis, implemented within
the SM, can accommodate most of the observed pattern of quark masses
and mixing, with the notable exception of the strength of CP violation,
measured by the rephasing invariant $I_{CP}\equiv|\mathrm{{Im}(V_{us}V_{cb}V_{cs}^{\ast}V_{ub}^{\ast})|}$.
This is a remarkable achievement, since the fit of experimental data
can be obtained even in the context USY ansätze \cite{usy-second}
\cite{usy-second0}, where the USY phases are entirely fixed by the
quark mass ratios, while the Cabibbo Kobayashi Maskawa mixing matrix
(CKM) is predicted without free parameters. Prior to the measurement
of the rephasing invariant angle $\gamma\equiv Arg(-V_{11}V_{23}V_{13}^{\ast}V_{21}^{\ast})^{CKM}$,
of the standard unitarity triangle, the fact that USY within the SM
cannot account for the observed strength of CP violation, was not
a major drawback. One could always assume that New Physics (NP) contributed
significantly to CP violation both in the Kaon and B-sectors, thus
allowing for a small value of $I_{CP}$. This is no longer viable
with the measurement of a large value of $\gamma$ \cite{pdg}, which
does not receive significant contributions from NP \cite{botella}.
However, NP can still give an important or even dominant contribution
\cite{aguilar} to $\chi\equiv Arg(-V_{23}V_{32}V_{22}^{\ast}V_{33}^{\ast})^{CKM}$,
which is small (of the order of a few percent) in the framework of
the SM \cite{aguilar}.

In this paper, we show that if one implements the USY hypothesis in
the framework of an extension of the SM, where one vector-like \cite{vectorlike}
quark is introduced, one may fully account for the observed pattern
of quark masses and mixing, including the size of $I_{CP}$. This
paper is organized as follows. In the next section, we analyse a specific
USY ansatz within the SM where the full CKM matrix is predicted as
a function of the quark mass ratios, with no free parameters\cite{usy-third}.
This analysis shows that the Ansatz is able to account for the pattern
of quark masses and mixing, with the exception of the strength of
CP violation. In section 3, we present an analytical study of a new
USY ansatz in the framework of an extension of the SM, where one isosinglet
vector-like quark is introduced. A numerical analysis of the same
Ansatz is presented in section 4. This analysis indicates that, in
the above extension of the SM, it is possible to account for the pattern
of quark masses and mixing, including the strength of CP violation.
Finally, we present our conclusions in section 5.

\section{USY predictions within the SM}

In this section, we illustrate the USY predictions within the SM,
in the framework of an Ansatz where the full CKM matrix is predicted
in terms of quark mass ratios, with no free parameters\cite{usy-third}.
Let us assume the following structure for the up and down quark mass
matrices 
\begin{equation}
M_{u,d}=c_{u,d}\left(\begin{array}{ccc}
1 & 1 & e^{i(\alpha-\beta)}\\
1 & 1 & e^{i(\alpha)}\\
e^{i(\alpha-\beta)} & e^{i(\alpha)} & e^{i(\alpha)}
\end{array}\right)_{u,d}\label{usy1}
\end{equation}
Since we have a total of three parameters for each sector, all quantities
of the CKM matrix are exact functions of the quark mass ratios. To
extract these, we proceed as follows. We define a squared mass matrix
$H\equiv\frac{1}{9c^{2}}MM^{\dagger}$ with trace normalized to $1$.
Then, the two remaining invariants of $H$, are expressed in terms
of the USY phases 
\begin{equation}
\begin{array}{l}
Det[H]\equiv\delta=\frac{m_{1}^{2}m_{2}^{2}m_{3}^{2}}{\left(m_{1}^{2}+m_{2}^{2}+m_{3}^{2}\right)^{3}}=\frac{4^{2}}{9^{3}}\sin^{4}(\frac{\beta}{2})\\
\\
\chi\lbrack H]\equiv\chi=\frac{m_{1}^{2}m_{2}^{2}+m_{1}m_{3}+m_{2}^{2}m_{3}^{2}}{\left(m_{1}^{2}+m_{2}^{2}+m_{3}^{2}\right)^{2}}=\frac{4}{9^{2}}\left[\sin^{2}(\frac{\alpha}{2})+4\sin^{2}(\frac{\beta}{2})+\sin^{2}(\frac{\alpha-2\beta}{2})+2\sin^{2}(\frac{\alpha-\beta}{2})\right]
\end{array}\label{inv1}
\end{equation}
Taking into account the observed hierarchy of quark masses, Eq. (\ref{inv1})
implies that both phases $\alpha,\beta$ have to be small. They can
be expressed as explicit functions of the $\chi$ and $\delta$ :
\begin{equation}
\begin{array}{l}
\sin^{2}(\frac{\beta}{2})=\frac{27}{4}\sqrt{\delta}\\
\\
\sin^{2}(\frac{\alpha-\beta}{2})=\frac{9^{2}}{4^{2}}\chi\left(\frac{1-2\frac{\sqrt{\delta}}{\chi}}{1-\frac{27}{4}\sqrt{\delta}}\right)
\end{array}\label{extract}
\end{equation}

The Ansatz of Eq. (\ref{usy1}) can be viewed as a specific perturbation
of the so called ''democratic ansatz'', which is obtained in the
limit $\alpha=\beta=0$. At a first stage of the perturbation, $\alpha\neq0$,
$\beta=0$ and the $u$, $d$ quarks are massless, so it corresponds
to the chiral limit. In fact, it has been shown that the USY structures
corresponding to the chiral limit can be classified into six exact
solutions \cite{usy-second0}, so Eq. (\ref{usy1}) with $\beta=0$
corresponds to one of these solutions \cite{usy-third}. At a second
stage of the perturbation, $\beta\neq0$ and the $u$, $d$ quarks
acquire a mass. From Eqs. (\ref{inv1}, \ref{extract}), one obtains
in leading order: 
\begin{equation}
\left|\alpha\right|=\frac{9}{2}\frac{m_{2}}{m_{3}}\quad;\quad\left|\beta\right|=3\sqrt{3}\frac{\sqrt{m_{1}m_{2}}}{m_{3}}\label{ab}
\end{equation}
 It is also useful to introduce the parameters $\lambda$ and $A$
defined by 
\begin{equation}
\lambda=\sqrt{\frac{m_{1}}{m_{2}}}\quad;\quad A=\frac{m_{2}^{2}}{m_{3}m_{1}}=O(1)\label{wolf}
\end{equation}
 It is clear that this choice of parameters was inspired by the Wolfenstein
parametrization of the CKM matrix. Using Eq. (\ref{wolf}), Eq. (\ref{ab})
translates into $\left|\alpha\right|=\frac{9}{2}\ A\lambda^{2}$,
$\left|\beta\right|=3\sqrt{3}\ A\lambda^{3}$.

\bigskip{}
The quark mass matrices $H_{u,d}\equiv\left(\frac{1}{9c^{2}}MM^{\dagger}\right)_{u,d}$
are diagonalized by: 
\begin{equation}
V_{u}^{\dagger}\ H_{u}\ V_{u}=D_{u}^{2}\quad;\quad V_{d}^{\dagger}\ H_{d}\ V_{d}=D_{d}^{2}\label{diag}
\end{equation}
The unitary matrices $V_{u,d}$ can be written as: 
\begin{equation}
V_{u,d}=F\cdot W_{u,d}\quad;\quad F=\left[\begin{array}{ccc}
\frac{1}{\sqrt{2}} & \frac{-1}{\sqrt{6}} & \frac{1}{\sqrt{3}}\\
\frac{-1}{\sqrt{2}} & \frac{-1}{\sqrt{6}} & \frac{1}{\sqrt{3}}\\
0 & \frac{2}{\sqrt{6}} & \frac{1}{\sqrt{3}}
\end{array}\right]\label{vd}
\end{equation}
The absolute values of the elements of the matrices $W_{u,d}$ are
close to those of the identity and can be expressed in terms of the
parameters $\lambda$ and $A$ as: 
\begin{equation}
\begin{array}{l}
\left\vert W_{12}\right\vert =\lambda\ \left(1+...\right)\\
\\
|W_{23}|=\sqrt{2}\ A\lambda^{2}\ \left(1+...\right)\\
\\
|W_{13}|=\frac{1}{\sqrt{2}}\ A\lambda^{3}\ \left(1+...\right)
\end{array}\label{v-simple}
\end{equation}
Since we take the same flavour structure for the down and up quark
sectors, \ the matrix $F$ cancels out in the computation of $V_{CKM}$
and one obtains: 
\begin{equation}
V_{CKM}=W_{u}^{\dagger}\ W_{d}\label{ckm1}
\end{equation}
This ansatz has the remarkable feature of giving $V_{CKM}$ in terms
of the quark mass ratios, with no free parameters. In leading order
one obtains: 
\begin{equation}
\begin{array}{l}
V_{12}^{CKM}=\sqrt{\frac{m_{d}}{m_{s}}}\left(1-\frac{3i}{4}\frac{m_{s}}{m_{b}}-i\sqrt{3}\sqrt{\frac{m_{d}}{m_{s}}}\frac{m_{s}}{m_{b}}\right)-\sqrt{\frac{m_{u}}{m_{c}}}\\
\\
V_{23}^{CKM}=\sqrt{2}\frac{m_{s}}{m_{b}}\left(1+\frac{\sqrt{3}}{2}\sqrt{\frac{m_{d}}{m_{s}}}+\frac{3i}{4}\frac{m_{s}}{m_{b}}+i\frac{\sqrt{3}}{2}\sqrt{\frac{m_{d}}{m_{s}}}\frac{m_{s}}{m_{b}}\right)\\
\\
V_{13}^{CKM}=\frac{-1}{\sqrt{2}}\frac{m_{s}}{m_{b}}\left(\sqrt{\frac{m_{d}}{m_{s}}}+2\sqrt{\frac{m_{u}}{m_{c}}}\left(1+\frac{3i}{4}\frac{m_{s}}{m_{b}}\right)-i\frac{\sqrt{3}}{2}\frac{m_{d}}{m_{s}}\frac{m_{s}}{m_{b}}\right)\\
\\
V_{22}^{CKM}=1-\frac{m_{d}}{2m_{s}}-\frac{m_{u}}{2m_{c}}+\sqrt{\frac{m_{d}}{m_{s}}}\sqrt{\frac{m_{u}}{m_{c}}}+\frac{3}{8}\left(\frac{m_{d}}{m_{s}}\right)^{2}-\left(\frac{m_{s}}{m_{b}}\right)^{2}
\end{array}\label{ckm2}
\end{equation}
where the signs in front of each of the mass ratio depend on the relative
sign of the initial USY phases{%
\footnote{Althougth there are other terms of the same order in each of these
$V_{CKM}$ elements, we only give here the terms which contribute
to $I_{CP}$ in lowest order. Up to this order, one may take $V_{22}^{CKM}=1$.
In addition, one may even vary the sign of each mass ratio $\pm\sqrt{\frac{m_{d}}{m_{s}}}$,
$\pm\frac{m_{s}}{m_{b}}$, $\pm\sqrt{\frac{m_{u}}{m_{c}}}$, depending
on the sign of the associated USY complex phase. However, despite
this extra freedom, significant cancelations still occur in the leading
order terms of $I_{CP}$, which remains small.%
}}. These results for the USY ansatz are very encouraging as they
lead to good phenomenological results for $V_{CKM}$ as a function
of the quark mass ratios. One can see that in a first order approximation,
the following sum rule holds in leading order: 
\begin{equation}
\left\vert \frac{V_{13}^{CKM}}{V_{23}^{CKM}}\right\vert =\frac{1}{2}\left\vert V_{12}^{CKM}\right\vert \label{sum}
\end{equation}
which is in agreement with experiment.

Having $V_{CKM}$ as a function of the quark mass ratios, one can
also compute $I_{CP}\equiv\mathrm{{Im}(}$ $V_{12}V_{23}V_{22}^{\ast}V_{13}^{\ast})^{CKM}$
as a function of the masses. The leading order term that one finds
is 
\begin{equation}
\left\vert I_{CP}\right\vert =\frac{3\sqrt{3}}{8}\left(\left\vert \left(\frac{m_{d}}{m_{s}}\right)\pm2\sqrt{3\frac{m_{u}}{m_{c}}}\right\vert \right)\sqrt{\frac{m_{d}}{m_{s}}}\left(\frac{m_{s}}{m_{b}}\right)^{3}\label{jcp}
\end{equation}
where again the sign depends on the relative sign of the initial USY
phases. However, Eq. (\ref{jcp}) illustrates the weak point of the
USY hypothesis. From Eq. (\ref{jcp}), one concludes that $\left\vert I_{CP}\right\vert \lesssim o(\lambda^{9})\approx5\times10^{-7}$
which is too small to account for the experimental value $\left\vert I_{CP}\right\vert \cong3\times10^{-5}$.
The point is that this USY ansatz predicts correctly the values of
$\left\vert V_{12}^{CKM}\right\vert $, $\left\vert V_{23}^{CKM}\right\vert $,
$\left\vert V_{13}^{CKM}\right\vert $, $\left\vert V_{22}^{CKM}\right\vert $,
but predicts a small value for $\gamma\equiv Arg(-V_{11}V_{23}V_{13}^{\ast}V_{21}^{\ast})^{CKM}$,
which in turn leads to a small value for $I_{CP}$. The above statement
may seem contradictory, since it is well known \cite{lavoura} that
from the precise knowledge of four independent moduli of $V_{CKM}$,
one can extract $\left\vert I_{CP}\right\vert $ using $3\times3$
unitarity. The contradiction is only apparent. The point is that a
reliable extraction of $\left\vert I_{CP}\right\vert $ from $\left\vert V_{12}^{CKM}\right\vert $,
$\left\vert V_{23}^{CKM}\right\vert $, $\left\vert V_{13}^{CKM}\right\vert $,
$\left\vert V_{22}^{CKM}\right\vert $ (or from the moduli of the
first two lines of $V_{CKM}$) requires a totally unfeasible precision
in the knowledge of the above four moduli. A simple way of understanding
this, is by realizing that the above extraction of $I_{CP}$ would
be equivalent to deriving the area of the squashed unitarity triangle,
corresponding to orthogonality of the first two lines of $V_{CKM}$,
from the knowledge of its sides.

In general, it is not possible to generate a correct value of $I_{CP}$
within the USY framework implemented in the framework of the SM. Indeed,
a numerical analysis reveals that, for all USY cases in the SM, not
restricted to the particular form of Eq. (\ref{usy1}), one is lead
to $\left\vert I_{CP}\right\vert <10^{-6}$, which is too small compared
to $\left\vert I_{CP}\right\vert _{\exp}$.

\section{USY with one extra vector-like quark}

\subsection{The Model}

The previous analysis provides motivation to implement the USY hypothesis
in the framework of a minimal extension of the SM, consisting of the
addition of the following fields to the SM spectrum:

- A vector-like down-type (i.e. $Q=-1/3$) quark $D$ added to the
3 generation left handed and right handed quark fields $Q_{L_{i}}$,
$d_{R_{i}}$, $u_{R_{i}}$ ;

- A complex scalar Higgs singlet $S$, singlet under $SU(2)\times U(1)$.

In addition, to simplify the analysis, we impose a $Z_{2}$ symmetry
under which all the fields of the SM transform trivially, while the
new fields $D_{L}$, $D_{R}$ and $S$ are odd. With this symmetry,
mass terms of the form $\overline{Q_{L_{i}}}\Phi\, D_{R}$ (i.e. the
coupling to the SM Higgs) are absent. The Yukawa couplings and mass
terms of $D$ are given by: $\overline{D}_{L}\ (f^{i}S+\widetilde{f}^{i}S^{*})\ d_{R_{i}}+M\ \overline{D}_{L}\ D_{R}$.
Upon spontaneous symmetry breaking, the Higgs bosons acquire VEV's:
$<\Phi>=v$, $<S>=V_{S}$. It is clear that there are two scales in
the model:

-- the scale $v$ of electroweak $SU(2)\times U(1)$ breaking and

-- the scale $V_{S}=<S>$, which can be much larger ($|V_{S}|\gg v$),
since $<S>$ does not break the $SU(2)\times U(1)$ gauge symmetry.
The same applies to $M$, which can be much larger than $v$, since
the mass term $\overline{D}_{L}\ D_{R}$ is a $SU(2)\times U(1)$
invariant.

We will assume a USY structure for the Yukawa couplings connecting
standard quarks, so the up quark mass matrix has the form of Eq. (\ref{inv1}),
while the down quark mass matrix is given by: 
\begin{equation}
\mathcal{M}_{d}=c_{d}\left(\begin{array}{cccc}
1 & 1 & e^{i(\alpha_{d}-\beta_{d})} & 0\\
1 & 1 & e^{i\alpha_{d}} & 0\\
e^{i(\alpha_{d}-\beta_{d})} & e^{i\alpha_{d}} & e^{i\alpha_{d}} & 0\\
p_{1} & p_{2} & p_{3} & \mu_{o}
\end{array}\right)\label{m4}
\end{equation}
where the $p_{i}$ are complex numbers given by $\ p_{i}=\frac{1}{c_{d}}(f^{i}V_{S}+\widetilde{f}^{i}V_{S}^{\ast}\ )$
and $\ \mu_{o}=\frac{1}{c_{d}}M$. In view of the above discussion,
it is natural to assume that $O\left(|p_{i}|\right)=O(\mu_{o})\gg1$.

\subsection{The effective down quark mass matrix}

Next we evaluate how the presence of the vector-like quark $D$ affects
the effective $3\times3$ down quark mass matrix connecting the standard
quarks. Note that, it is this effective down quark mass matrix $H_{eff}^{d}$
which, together with the up quark mass matrix, that generates the
$3\times3$ CKM matrix connecting standard quarks.

In order to pursue our goal, we explicitly compute the effective $3\times3$
down quark mass matrix, resulting from Eq. (\ref{m4}) by integrating
out the fourth heavy down quark{%
\footnote{For convinience and instead of the dimensionfull effective matrix
$\mathbf{H}_{eff}^{d}$, we shall work with a dimensionless matrix
$H_{eff}^{d}$, defined as $H_{eff}^{d}\equiv\frac{1}{9c_{d}^{2}}\mathbf{H}_{eff}^{d}$.%
}}. We explain exactly how the extra vector-like quark modifies the
pure USY hypothesis, and thus, gives a significant contribution to
CP violation. We obtain in leading order: 
\begin{equation}
H_{eff}^{d}=m\ m^{\dagger}-m\ \frac{P\ P^{\dagger}}{\mu^{2}}\ m^{\dagger}\label{heff0}
\end{equation}
with 
\begin{equation}
\begin{array}{l}
m=\frac{1}{3}\left(\begin{array}{ccc}
1 & 1 & e^{i(\alpha-\beta)}\\
1 & 1 & e^{i(\alpha)}\\
e^{i(\alpha-\beta)} & e^{i(\alpha)} & e^{i(\alpha)}
\end{array}\right)\quad;\quad P=\left(\begin{array}{c}
p_{1}^{\ast}\ \\
p_{2}^{\ast}\ \\
p_{3}^{\ast}
\end{array}\right)\\
\mu^{2}=|p_{1}|^{2}+|p_{2}|^{2}+|p_{3}|^{2}+\mu_{o}^{2}
\end{array}\label{heff}
\end{equation}
where we have dropped the down-quark subscript $d$ and for convenience
have normalized the trace of $m\ m^{\dagger}$ such that $Tr(m\ m^{\dagger})=1$.
Note that the two terms of $H_{eff}^{d}$ in Eq. (\ref{heff0}) are
of the same order. This is a crucial point, since it makes possible
to have a significant contribution to CP violation arising from the
vector-like quark $D$. In particular, the contribution of the vector-like
quark to $H_{eff}^{d}$ does not decouple in the limit of large vector-like
quark-mass, provided $V_{S}$ and $M$ in Eq. (\ref{m4}) are of the
same order of magnitude. This is to be contrasted with the effect
of the vector-like quark on the tree-level flavour changing neutral
currents (FCNC) which, although present in this class of models, are
naturally suppressed by the ratio $\overline{m}^{2}/\overline{M}^{2}$,
where $\overline{m}$ is the standard quark mass and $\overline{M}$
denotes the mass of the heavy vector-like quark. We shall return to
this question in the sequel.

In addition (for our case), one can prove that $H_{eff}^{d}$ yields
the same results as the full down $4\times4$ square mass matrix $\mathcal{M}_{d}\mathcal{M}_{d}^{\dagger}$
up to corrections of the order of $O(\left({\frac{m_{b}}{ \overline{M} } }\right)^{2})$,
with $\overline{M}$ denoting the mass of the extra vector-like down quark.
We use $H_{eff}^{d}$ in order to have an analytical understanding
of the effects of the vector-like quark. However, in our numerical
analysis, we shall include the calculation with the full $4\times4$
square mass matrix $\mathcal{M}_{d}\mathcal{M}_{d}^{\dagger}$, its
mass eigenvalues and diagonalization matrix.

\subsubsection{FCNC}

An important effect of the presence of extra vector-like quarks is
the occurrence at tree level of flavour changing neutral currents
(FCNC) and their associated electroweak precision measurements (EWPM),
which place severe restrictions on this class of models. For our case,
with one extra vector-like down quark, FCNC only occur in the coupling
of the $Z$ with the down quarks: 
\begin{equation}
-\frac{g}{2\cos(\theta_{W})}\overline{d}_{L}^{\alpha}\ \gamma^{\mu}\ \mathcal{U}_{\alpha\beta}\ d_{L}^{\beta}\ Z_{\mu}\label{change}
\end{equation}
where $\mathcal{U}$ is a $4\times4$ matrix computed from $\mathcal{V}$
the full $3\times4$ CKM matrix{%
\footnote{In the case of the SM, the CKM matrix $\mathcal{V}$ coincides with
an ordinary $3\times3$ unitary matrix $V$, and then, $\mathcal{U}$
is the $3\times3$ identity matrix, so there are no FCNC's at tree
level.%
}}, $\mathcal{U}=\mathcal{V}^{\dagger}\ \mathcal{V}$. The CKM matrix
$\mathcal{V}$ is obtained from $\mathcal{V=}V_{u}^{\dagger}\ \mathcal{K}_{d}$
, where $V_{u}$ is the unitary matrix which diagonalizes the $3\times3$
(squared) up quark mass matrix $H_{u}$, and $\mathcal{K}_{d}$ is
the $3\times4$ part of the unitary matrix $\mathcal{W}_{d}=\binom{\mathcal{K}_{d}}{\beta_{d}}$
which diagonalizes the full $4\times4$ (squared) down quark mass
matrix $\mathcal{M}_{d}\mathcal{M}_{d}^{\dagger}$ in Eq. (\ref{m4}).
Since $V_{u}$ is unitary, one has $\mathcal{V=}\mathcal{K}_{d}^{\dagger}\ \mathcal{K}_{d}$.
The CKM matrix is not unitary, but its effective deviation from unitarity
is very small, being naturally suppressed by the ratio ${\overline{m}}^{2}/\overline{M}^{2}$,
where $\overline{m}$ is the standard quark mass and $\overline{M}$ denotes
the mass of the heavy vector-like quark.

The most stringent EWPM is the branching ratio of $K^{+}\rightarrow\pi^{+}\nu\overline{\nu}$,
which places a severe limit on the FCNC coupling $\mathcal{U}_{sd}$.
This branching ratio, and other quantities sensitive to FCNC, which
depend on the $\mathcal{U}_{ij}$, have been computed for models with
vector-like down quarks \cite{Knunu}. In our case, we typically obtain
values $\left\vert \mathcal{U}_{sd}\right\vert ,\left\vert \mathcal{U}_{ds}\right\vert \leq\times10^{-9}$,
$\left\vert \mathcal{U}_{bd}\right\vert ,\left\vert \mathcal{U}_{db}\right\vert \leq\times10^{-8}$,
$\left\vert \mathcal{U}_{bs}\right\vert ,\left\vert \mathcal{U}_{sb}\right\vert \leq\times10^{-6}.$
For these values, we find no significant effect on the associated
EWPM.

\subsection{Extra CP violation from the vector-like quark}

In order to explicitly show the\ effect of the extra vector-like
quark, we choose as an example: $(\ p_{1}\ ,\ p_{2}\ ,\ p_{3}\ ,\ \mu\ )=(\ ik\ ,\ ik\ ,-k\ ,\ k\ )$.
For this case, we see that the dependence on $k$ cancels out and
the extra term $\frac{P\ P^{\dagger}}{\mu^{2}}$ in Eq. (\ref{heff0})
reduces to a simple expression: 
\begin{equation}
\frac{P\ P^{\dagger}}{\mu^{2}}=\frac{1}{4}\ \ d_{i}\ \Delta\ d_{i}^{*}\quad\quad;\quad\quad d_{i}=diag(i\ ,i\ ,1)\quad\quad;\quad\quad\Delta=\left(\begin{array}{ccc}
1 & 1 & 1\\
1 & 1 & 1\\
1 & 1 & 1
\end{array}\right)\label{extra}
\end{equation}
 and therefore, 
\begin{equation}
H_{eff}^{d}=m\ m^{\dagger}-\frac{1}{4}\ m\ d_{i}\ \Delta\ d_{i}^{*}\ m^{\dagger}\label{case1}
\end{equation}

The diagonalization of this effective matrix is done in two steps.
First, we use the result obtained in Eqs. (\ref{wolf}, \ref{vd},
\ref{v-simple}), where we have expressed the diagonalization matrix
$V=F\cdot W$ of $m\ m^{\dagger}$ as the product of a fixed unitary
matrix $F$ and a unitary matrix $W$, whose absolute value is close
to the identity and which can be parametrized à la Wolfenstein. The
matrix $m$ is symmetric, thus we have $V^{\dagger}m\ V^{\ast}=d$,
where $d$ is diagonal and real. All parameters, including the USY
phases, the eigenvalues of $m\ m^{\dagger}$ and $W$, are given as
expansions in a small parameter $\overline{\lambda}$, and as we chose
$Tr(m\ m^{\dagger})=Tr(d^{2})=1$, we have for the leading order terms:
$d=diag(\ \overline{A}\overline{\lambda}^{4}\ ,\ \overline{A}\overline{\lambda}^{2}\ ,\ 1)$.
Therefore, partially diagonalizing $H_{eff}^{d}$ we obtain 
\begin{equation}
H_{eff}^{d}\longrightarrow\widetilde{H}_{eff}^{d}=V^{\dagger}H_{eff}^{d}\ V=d^{2}-\frac{1}{4}\ d\ W^{T}\ F^{T}\ d_{i}\ \Delta\ d_{i}^{\ast}\ F\ W^{\ast}\ d\label{part}
\end{equation}
Note that we chose different symbols $\overline{\lambda}$ , $\overline{A}$
instead of the original $\lambda$, $A$. This is necessary because
the eigenvalues of $m\ m^{\dagger}$ do not coincide with those of
$H_{eff}^{d}$. Only the eigenvalues of $H_{eff}^{d}$ relate to the
real physical masses. For simplicity, let us for now assume that $W={1\>\!\!\!\mathrm{I}}$,
then in leading order we have 
\begin{equation}
\begin{array}{l}
\widetilde{H}_{eff}^{d}=d^{2}-\frac{1}{4}\ d\ \left(\begin{array}{rrr}
0 & 0 & 0\\
0 & \frac{4}{3} & \frac{-\sqrt{2}(1+3i)}{3}\\
0 & \frac{-\sqrt{2}(1-3i)}{3} & \frac{5}{3}
\end{array}\right)\ \ d=\\
\\
=\left(\begin{array}{rrr}
\overline{A}^{2}\overline{\lambda}^{8} & 0 & 0\\
0 & \frac{2}{3}\ \overline{A}^{2}\overline{\lambda}^{4} & \frac{(1+3i)}{6\sqrt{2}}\overline{A}\overline{\lambda}^{2}\\
0 & \frac{(1-3i)}{6\sqrt{2}}\overline{A}\overline{\lambda}^{2} & \frac{7}{12}
\end{array}\right)
\end{array}\label{part1}
\end{equation}
We see that to complete the total diagonalization of $H_{eff}^{d}$,$\ $we
have to diagonalize an extra $2\times2$ Hermitian matrix. This will
give an important contribution to $V_{CKM}$, since the extra term
corresponds to two modifications:

-- first, further diagonalization implies the transformation of the
matrix $\widetilde{H}_{eff}^{d}$ in Eq. (\ref{part1}) with an additional
diagonal unitary matrix $K=diag(1,1,\frac{(1-3i)}{\sqrt{10}})$ and
with an additional orthogonal matrix $O_{23}$ such that, in leading
order 
\begin{equation}
D_{eff}^{2}=O_{23}^{T}\ K^{*}\ \widetilde{H}\ _{eff}^{d}\ K\ O_{23}=\left(\begin{array}{rrr}
\overline{A}^{2}\overline{\lambda}^{8} & 0 & 0\\
0 & \frac{3}{7}\ \overline{A}^{2}\overline{\lambda}^{4} & 0\\
0 & 0 & \frac{7}{12}
\end{array}\right)\label{deff}
\end{equation}
 where $O_{23}$ is given in leading order by 
\begin{equation}
O_{23}=\left(\begin{array}{lll}
1 & 0 & 0\\
0 & 1 & \frac{2\sqrt{5}}{7}\overline{A}\overline{\lambda}^{2}\\
0 & -\frac{2\sqrt{5}}{7}\overline{A}\overline{\lambda}^{2} & 1
\end{array}\right)\label{o23}
\end{equation}

-- and second, since the trace in Eq. (\ref{deff}) of the total effective
matrix is no longer normalized to $1$, this implies that the parameters
$\overline{\lambda}$ and $\overline{A}$ are slightly different from
the original $\lambda$, $A$. We stress again, that this results
from the fact that the eigenvalues of $m\ m^{\dagger}$ are not the
same as the eigenvalues of $H_{eff}^{d}$. From Eq. (\ref{deff}),
it easy to conclude that, for this effective mass matrix, we now have
in leading order 
\begin{equation}
\overline{\lambda}=\left(\frac{3}{7}\right)^{\frac{1}{4}}\sqrt{\frac{m_{1}}{m_{2}}}\quad\quad;\quad\quad\overline{A}=\frac{7\sqrt{7}}{6\sqrt{3}}\frac{m_{2}^{2}}{m_{3}m_{1}}\label{Ala}
\end{equation}
which will give a significant contribution to $V_{CKM}$. The masses,
of course, refer to the down sector. The angle of the extra diagonalization
matrix $O_{23}$ is, in leading order, equal to $\frac{\sqrt{5}}{3}\frac{m_{2}}{m_{3}}$
and of the same order as the corresponding angle in the original diagonalizing
matrix $W$.

In conclusion, when computing the new diagonalization matrix, from
the effective mass matrix $H_{eff}^{d}$ , not only do we have to
include the extra unitary matrices $K$ and $O_{23}$, but we have
also to substitute the new expressions for $\overline{\lambda}$ and
$\overline{A}$ by the original $\lambda$, $A$ in Eq. (\ref{vd}).
Thus, 
\begin{equation}
V_{eff}^{d}=F\ \overline{W}\ K\ O_{23}\label{veff}
\end{equation}
where $\overline{W}=W(\overline{\lambda},\overline{A})$. Obviously,
from the form of $V_{eff}^{d}$, it is clear that CP violation gets
an extra contribution as it now includes the complex matrix $K$,
which can not be factorized out, i.e. included in redefinitions of
the quark fields. It is now easy to see that the leading order term
of $I_{CP}$ is just 
\begin{equation}
\begin{array}{l}
\left\vert I_{CP}\right\vert =\left(\frac{2\sqrt{5}}{7}\overline{A}\overline{\lambda}^{2}\right)\left(\frac{2}{\sqrt{5}}\right)\ \left\vert \overline{W}_{12}\right\vert \ \left(\left\vert \overline{W}_{12}\overline{W}_{23}\right\vert +\left\vert \overline{W}_{13}\right\vert \right)=\\
\\
=\left(\frac{2\sqrt{5}}{7}\overline{A}\overline{\lambda}^{2}\right)\left(\frac{2}{\sqrt{5}}\right)\ \overline{\lambda}\ \left(\frac{3}{2}\sqrt{2}\overline{A}\overline{\lambda}^{3}\right)=\\
\\
=\frac{6\sqrt{2}}{7}\overline{A}^{2}\overline{\lambda}^{6}=\sqrt{\frac{7}{6}}\left(\frac{m_{d}}{m_{s}}\right)\left(\frac{m_{s}}{m_{b}}\right)^{2}
\end{array}\label{jcp-leading}
\end{equation}
where it is sufficient to take the first order terms of $\overline{W}$
in~Eq.(\ref{v-simple}) and substitute the original $\lambda$ and
$A$ by the new $\overline{\lambda}$, $\overline{A}$ in Eq.(\ref{Ala}).
The factor $\frac{2\sqrt{5}}{7}\overline{A}\overline{\lambda}^{2}$stands
for the angle of the extra diagonalization matrix $O_{23}$ and the
factor $\frac{2}{\sqrt{5}}$ comes from the contribution from the
extra unitary matrix $K$ together with the phases{%
\footnote{For the sake of the argument, we have taken until now as an example
$W={1\>\!\!\!\mathrm{I}}$, however it turns out that, it is the absolute
values of the elements of the matrices $W$ that are close to those
of the identity. In fact, $W$ itself differs from the identity by
a diagonal unitary matrix. To be correct, $W$ is very close to $diag(-i,1,ie^{i\frac{\pi}{4}})$.%
}} of $\overline{W}$. We see now that $I_{CP}$ has become much larger:
$\left\vert I_{CP}\right\vert \approx o(\lambda^{6})$ which is very
near to its experimental value $3\times10^{-5}$.

\section{Exact numerical results}

\subsection{Numerical examples}

Next, we present exact numerical results. As previously mentioned,
the up quark mass matrix is taken to be a $3\times3$ pure USY matrix,
given in Eq. (\ref{usy1}), while the down quark matrix is a $4\times4$
matrix as in Eq. (\ref{m4}). We present 3 kinds of cases and results
to be compared:
\begin{itemize}
\item [(i)] We give a numerical example with the full $3\times4$ CKM
matrix and other physical quantities using the full $4\times4$ down-type
quark mass matrix.
\item [(ii)] We calculate the same output using the $3\times3$ effective
down-type quark mass matrix, according to Eq. (\ref{heff0}). 
\end{itemize}
\textbf{Input:}

\begin{center}
\begin{equation}
\begin{tabular}{lllcl}
 \ensuremath{c_{u}=57.334\ \mathrm{{GeV}\,}} &   &  \ensuremath{\alpha_{u}=0.0168} &  \hspace{5mm}  &  \ensuremath{p_{1}=90+180\ i}\\
\ensuremath{c_{d}=1.163\ \mathrm{{GeV}\,}} &   &  \ensuremath{\beta_{u}=0.00088} &  \hspace{5mm}  &  \ensuremath{p_{2}=-20+125\ i}\\
 &   &  \ensuremath{\alpha_{d}=0.108} &  \hspace{5mm}  &  \ensuremath{p_{3}=-365+20i}\\
 &   &  \ensuremath{\beta_{d}=-0.028} &  \hspace{5mm}  &  \ensuremath{\mu_{o}=190}
\end{tabular}\label{input}
\end{equation}
 
\par\end{center}

\textbf{Output (at the }$M_{Z}$\textbf{\ scale):}
\begin{itemize}
\item [(i)] Output using the $4\times4$ down-type mass matrix: 
\begin{equation}
\begin{array}{l}
(m_{u},m_{c},m_{t})=(0.00139,\,0.610,\,172)\ \mathrm{{GeV}\,,}\\
\\
(m_{d},m_{s},m_{b},m_{D})=(0.00321,\,0.0529,\,2.9,\,553)\ \mathrm{{GeV}\,,}\\
\\
|\mathcal{V}_{CKM}|=\left(\begin{array}{cccc}
0.97436 & 0.22497 & 0.003606 & 0.0000352\\
0.22480 & 0.97351 & 0.04163 & 0.000137\\
0.009371 & 0.04072 & 0.99912 & 0.00350
\end{array}\right)\,,\\
\\
\rho\equiv|\mathcal{V}_{13}|/|\mathcal{V}_{23}|=0.0866\,,\qquad|I_{CP}|=3.2\times10^{-5}\,,\\
\\
\sin(2\beta)=0.685\,,\qquad\gamma=79.2^{o}\,.\\
\\
\mathrm{FCNC}:\,\left\vert \mathcal{U}_{sd}\right\vert ,\left\vert \mathcal{U}_{ds}\right\vert =5.3\times10^{-10},\left\vert \mathcal{U}_{bd}\right\vert ,\left\vert \mathcal{U}_{db}\right\vert =8.4\times10^{-9},\left\vert \mathcal{U}_{bs}\right\vert ,\left\vert \mathcal{U}_{sb}\right\vert =7.2\times10^{-7}
\end{array}\label{out1}
\end{equation}
\\

\item [(ii)] Output using the $3\times3$ effective down-type mass matrix
approximation of Eq. (\ref{heff}): 
\begin{equation}
\begin{array}{l}
(m_{u},m_{c},m_{t})=(0.00139,\,0.610,\,172)\mathrm{{GeV}\,,}\\
\\
(m_{d},m_{s},m_{b})=(0.00321,\,0.0529,\,2.9)\mathrm{{GeV}\,,}\\
\\
|V_{CKM}|=\left(\begin{array}{ccc}
0.97436 & 0.22497 & 0.003606\\
0.22480 & 0.97351 & 0.04163\\
0.009371 & 0.04072 & 0.99913
\end{array}\right)\,,\\
\\
\rho\equiv|V_{13}|/|V_{23}|=0.0866\,,\qquad|I_{CP}|=3.2\times10^{-5}\,,\\
\\
\sin(2\beta)=0.685\,,\qquad\gamma=79.2^{o}\,.
\end{array}\label{out2}
\end{equation}
 
\end{itemize}
The exact numerical results of Eq. (\ref{out1}) show that a good
fit of the quark masses and mixing, including the strength of CP violation
can be obtained in the USY framework with the addition of one down-type
singlet quark. Comparison of Eqs. (\ref{out1}, \ref{out2}) shows
that the use of the effective down-type mass matrix is an excellent
approximation, as anticipated. Furthermore, the off-diagonal elements
$\mathcal{U}_{ij}$ of the FCNC mixing matrix are very small. Using
the expressions given in \cite{Knunu}, and these elements we have
computed e.g. the branching ratio of $K^{+}\rightarrow\pi^{+}\nu\overline{\nu}$.
We have found that our FCNC mixing matrix elements have no significant
influence on the associated EWPM's.

It is important to verify that a good fit of quark masses and mixing,
including the strength of CP violation is obtained in a non-singular
region of parameter space, so that the results are stable. This was
verified by doing a numerical scan of the parameter space. We chose
only a region of the full parameter space where we expect to obtain
good results. The input parameter space consists of the parameters
$(\alpha_{u},\,\beta_{u}\,,\,\alpha_{d},\,\beta_{d},\, ip_{1},\, ip_{2},-\, p_{3},\,\mu_{o})$.
For each of these, we chose initial and final values and a step. We
maintained the top and bottom mass at fixed values of $171$ and $2.9$
GeV's at $M_{Z}.$ We obtained a dense set of points around the values
presented in the above example, all of them corresponding to values
of masses and mixing allowed by the experimental data.

\section{Conclusions}

We have shown that a good fit of quark masses and mixing, including
the strength of CP violation can be obtained in an extension of the
Standard Model with one down-type singlet quark and universal strength
of Yukawa couplings connecting standard quarks. The results were obtained
both through an analytical approximate diagonalization of the quark
mass matrices and through an exact numerical evaluation of quark masses
and mixing. The crucial point is that, in the presence of the vector-like
quark singlet, the effective $3\times3$ quark mass matrix receives
in leading approximation two contributions, one arising from the standard
quark mass matrices and another one arising from quark mass terms
involving the vector-like quark and its mixing with standard quarks.
This new contribution to the effective mass matrix is not suppressed
by the ratio of the standard quark masses and singlet quark masses.
It is this new contribution to the effective quark mass matrix which
increases the strength of CP violation, measured by the rephasing
invariant $I_{CP}$. It is worth emphasizing that for a broad class
of flavour models, the presence of heavy vector-like quarks can have
an important effect on the structure of low-energy quark masses and
mixing. This is an important point, in view of the significant number
of flavour models where vector-like quarks are introduced \cite{vector-R}.

At this stage the following comment is in order. We have introduced
only one isosinglet vector-like quark. In what concerns the effective
quark mixing and CP violation in the quark sector, this is equivalent
to having more than one isosinglet quark, with only one of these new
quarks mixing significantly with the standard quarks. Still, we have
not imposed the strict USY principle in the mixing terms like $f^{i}\ \overline{D}_{L}\ S\ d_{R_{i}}$
{%
\footnote{Even so, it is obvious that imposing the USY structure separately
on the $f^{i}$ and $\widetilde{f}^{i}$of the full interaction term
$\overline{D}_{L}\ (f^{i}S+\widetilde{f}^{i}S^{\ast})\ d_{R_{i}}$
will result in different absolute values of the respective mass matrix
elements.%
}}. Naively, this could seem unnatural, but this is not the case.
Note that for simplicity, we have introduced a $Z_{2}$ symmetry which
eliminates some Yukawa couplings, but this is not a crucial ingredient
of the model. Without this $Z_{2}$ symmetry, one would have mass
terms like $M^{i}\ \overline{D}_{L}\ d_{R_{i}}$ which are $SU(2)\times U(1)$\ invariant.
Consider now that there is a symmetry which would lead to USY in the
Yukawa terms connecting the standard quarks. That could be softly
broken by the mass terms $\overline{D}_{L}\ d_{R_{i}}$ without affecting
the naturalness of USY in the standard quark sector. On the other
hand, if one allows more than one isosinglet quark with significant
mixing with standard quarks, one may obtain sufficient CP violation
\cite{Higuchi:2006sx} even with quark mass terms $M^{i}\ \overline{D}_{L}\ d_{R_{i}}$
of equal size. The key point of this paper is showing that sufficient
CP violation is generated in the USY framework even if only one isosinglet
quark mixes significantly with standard quarks. Such a scenario is
entirely plausible if the spectrum of the isosinglet quarks is hierarchical,
where one expects that only the lightest isosinglet quark mixes significantly
with standard quarks.

It is remarkable that the USY Ansatz implemented in a simple extension
of the SM can accommodate the pattern of quark masses and mixing,
including the strength of CP violation. The next step would be to
find a framework where the USY structure for the Yukawa couplings
would result from a fundamental symmetry, eventually implemented at
a higher energy scale. Efforts to derive the USY structure from higher
dimension theories have been considered in the literature \cite{usy-dimensions}.
It is worth pointing out that, in higher dimensional theories, the
Kaluza Klein modes of some fermion fields behave effectively as vectorlike
fermions.

\section*{Acknowledgements}

This work was partially supported by Fundação para a Ciência e a Tecnologia
(FCT, Portugal) through the projects CERN/FP/109305/2009, PTDC/FIS/098188/2008
and CFTP-FCT Unit 777 which are partially funded through POCTI (FEDER)
and by Marie Curie Initial Training Network \textquotedbl{}UNILHC\textquotedbl{}
PITN-GA-2009-237920.

\end{document}